\let\ti=\times
\let\pa=\partial
\let\De=\Delta
\let\al=\alpha

\let\ra=\rightarrow
\let\va=\varphi

\let\vare=\varepsilon
\let\la=\lambda

\let\disp=\displaystyle
\let\ti=\times

\let\om=\omega

\let\De=\Delta

\def\que#1#2{\displaystyle\frac{#1}{#2}}

\documentclass[pra,twocolumn]{revtex4}

\usepackage{amsfonts,amssymb,bm,graphicx}

\begin{document}

\title{Zero-Point Radiation and the Big Bang}

\author{R. Alvargonz\'alez and L. S. Soto}

\affiliation{Facultad de F\'{\i}sica,  Universidad Complutense,
28040 Madrid, Spain}

\date{\today}

\begin{abstract}
This paper develops a cosmological hypothesis based on the following
propositions:
\begin{enumerate}
\item Zero-point radiation derives from quantic fluctuations in space, and the
wavelength of its photons with the greatest energy is inversely proportional
to the curvature of space.
\item The Universe began as the breaking in of photons of extremely high
energy contained in the 3-dimensional surface: $w^2+x^2+y^2+z^2=R^2_i$, whose
radius has continued to expand at the speed of light since its origin at
$t=0$.
\item The wavelength of the photons is quantized and the quantum of wavelength
is invariable.
\end{enumerate}

These propositions imply that the value of the total energy of
the zero-point radiation in the Universe remains constant and
the condition $w^2 + x^2 + y^2 + z^2 = (R_i + ct)^2 = R_u^2$
determines that every point in our space is subject to a
tension whose intensity $i$ is proportional to the curvature
$1/R_u$. Any increase of $R_u$ implies a decrease in $i$ and
consequently an energy flow which translates into an expansive
force. Therefore, the Universe will expand indefinitely:
no Big Crunch is possible. If the initial radius of the
Universe $R_i$ has been smaller than the Schwarzschild radius,
$R_s$, which corresponds to the total mass of the Universe,
$M_u$, the generation of matter would have lasted for thousands
of millions of years. Generation of matter over short periods
would have required values for $R_i$ of thousands of millions
of light years.
\end{abstract}

\maketitle

\section{Introduction}

We must distinguish between the Universe, the material Universe and the
visible Universe. The radius of the Universe, $R_u$, measures $R_i+ct$, where
$t$ is the time elapsed since $t=0$, but we are unable to measure it. The
material Universe consists of elementary particles and cosmic objects.
Obviously, its radius, $R_m$, is shorter than $R_u$.

The expansion  of the Universe determines that everything which is very far
away from us, may recede at velocities equal or greater than $c$, which
implies the existence of a horizon of visibility. We can only observe
extremely luminous cosmic objects which recede at a speed near that of light.
The value of the angle $\va$ which defines the visible Universe is such than 1
radian $<\va$ radians $<\pi$ radians; $1/\va=v_m/c$, where $v_m$ is the
present rate of increase in the length of the radius of the material Universe.
Both $\va$ and $v_m$ are functions of the relation, $R_0/R_s$, between the
radius of the material Universe at the end of the generation of matter, $R_0$,
and the Schwarzschild radius, $R_s$.

\vskip 6pt
\begin{center}*\ \ \ *\ \ \ *\end{center}\vskip 16pt

In 1916, Nernst suggested that the quantic fluctuations of space must cause
an electromagnetic radiation which would therefore be inherent to space and,
consequently, have a spectrum which is relativistically invariant.

In 1958, Sparnaay found this radiation when he was measuring the Casimir
effect at temperatures close to absolute zero. He detected some radiation
which was independent of temperature, with a spectrum such that the
intensities of its flows are inversely proportional to the cubes of their
wavelengths, which is a necessary condition for the radiation to be
relativistically invariant [1], [2], [3]. In 1997, S. K. Lamoureux carried out
new measurements of the intensity of the energy flow of zero-point radiation,
using a very different method, and reached the same measurements as Sparnaay's
[12].

A function of spectral distribution which is inversely proportional to the
cubes of the wavelengths, implies a distribution of energies which is
proportional to the $4^{\rm th}$ power of the wavelengths, because the
energies of the photons are inversely proportional to the wavelengths. In
1969, Timothy H. Boyer, [7], [8], showed that the spectral density function
of zero-point radiation is:
$$f_\va(\la)=\que1{2\pi^2} \que1{(\la_*)^3},\eqno{(1)}$$
where $\la_*$ is the number giving the measurement of the wavelength $\la$.

This function produces the next, for the corresponding energies
$$E_\va(\la)=\que1{2\pi^2} \que{hc}\la \que1{(\la_*)^3}.\eqno{(2)}$$

For $\la\ra0$, $E_\va(\la)\ra\infty$. There must be, therefore, a threshold
for $\la$, which will be hereafter designated by the symbol $q_\la$.

\vskip 6pt
\begin{center}*\ \ \ *\ \ \ *\end{center}\vskip 16pt

To simplify the following arguments, it is convenient to use the $(e,m_e,c)$
system of units in which the basic units are the quantum of electric charge,
$e$, the mass of the electron, $m_e$, and the speed of light, $c$. In this
system the units of length and time are, respectively, $l_e=e^2/(m_ec^2)$ and
$t_e=e^2/(m_ec^3)$. The unit of length is equal to the classic radius of the
electron.

Zero-point radiation proceeds equally from all directions of space, and its
interactions with electrons could, therefore, play the role of the
``Poincar\`e tensions", preventing the electron from shattering as a result of
the repulsion of its charge against itself. For his to be the case, there
must operate the equation:
$$x^3=\que{4\pi^3}{3\al}(k_\la)^4(r_x)^4[B]_m;\eqno{(3)}$$
equation 17 in [4], where:
\begin{description}
\item[$x=$] measurement of the wavelength of the photons with the greatest
energy in zero-point radiation, expressed in $q_\la$ (quanta of wavelength).
\item[$k_\la=$] measurement of $l_e$, expressed in $q_\la$.
\item[$r_x=$] measurement of the radius of the electron, expressed in $l_e$.
\end{description}
$$[B]_m=\que7{48}B-\que{11}{50}B^2+\cdots+T_mB^m,$$
where:
$$B=\que{2\pi}\al\left(\que{k_\la}x\right)$$
$$T_m=(-1)^{m-1}$$
$$\left[\que1{m+1}+\que2{m+2}-\que3{m+3}-1-
\que{m(m-1)}6\right]\que1{m+3}$$

The hypothesis that zero-point radiation is also the effective cause of
gra\-vi\-ta\-tio\-nal attraction between two electrons leads to the equation:
$$x^3=\que{2\pi^2}{3\al}\que{(k_x)^2(r_x)^2[B]_m}{G_e}; \eqno{(4)}$$
equation 20 in [4].

In [4] it was also deduced that:
$$r_x=1l_e$$
$$k_\la=\left[\que1{2\pi G_e}\right]^{1/2};\qquad
G_e=\que1{2\pi(k_\la)^2}\eqno{(5)}$$
$q_\la=(2\pi\al)^{1/2}L_P$, where $L_P$ is the Planck length.
$$\left.\begin{array}{l}
k_x=8.143375\ti 10^{20}\\[+5pt]
x=5.257601\ti10^{27}\end{array}\right\}\eqno{(6)}$$

\section{Basic Principles of the Proposed Hypothesis}

The proposed hypothesis rests on the following basic principles:
\begin{enumerate}\itemsep=-2pt
\item The ``Big Bang"\ consisted of the appearance, at $t=0$, of a primal
space configured as the 3-dimensional surface $w^2+x^2+y^2+z^2=(R_i)^2$ of
radius $R_i$ light-years, whose zero-point radiation
was characterized by the fact that
its photons of greater energy possessed a wavelength of
$x_iq_\la=k_u(R_i/R_u)q_\la$. Within the primal space there would have
existed photons unconnected to that zero-point radiation.
\item After a lapse of $t$ years those photons which had not been transformed
into elementary particles would, after travelling in all directions,
have covered a
distance of $R_t=(R_i+t)$ light years, and the wavelength of the photons of
greatest energy in zero-point radiation would have increased to
$x_tq_\la=k_u(R_t/R_u)q_\la$, so that
the wavelengths of these photons would
always be directly proportional to the radius of the Universe and, therefore,
inversely proportional to its curvature. $k_u=\que{R_u}x=7.264351\ti10^{33}$.
\item  The quantum of wavelength, $q_\la$, is intrinsically invariant.
\end{enumerate}

\vskip 6pt
\begin{center}*\ \ \ *\ \ \ *\end{center}\vskip 16pt

The analysis of the proposed hypothesis requires us to consider how the
numbers $x,\ k_\la$ and $G_e$, as also the mass of the electron, have evolved,
according to its principles, from $t=0$ until the present, since any
impossibility or absurdity in that evolution would compel us to reject the
hypothesis. We do not need to extend our consideration to the gravitational
constant, $G=G_el_ec^2m_e^{-1}$, because the terms of the hypothesis imply the
invariance of this constant. In fact:
$$G_e=1/2\pi(k_\la)^2;\qquad l_e=k_\la q_\la,$$
and the product, $*_e=m_el_e$, of the mass of the electron and its
radius is a fundamental quantic threshold, no matter what values
of $m_x$ and $r_x$ are possessed by the mass and the radius of the
electron, the product $(m_xm_e)(r_xl_e)$ will always be equal to
$m_e  l_e=1*_e$. Keeping this in mind, we can write for the
gravitational constant,
$$G=\que1{2\pi(k_\la)^2}\que{l_e^2c^2}{m_el_e}=
\que{(k_\la)^2(q_\la)^2c^2}{2\pi(k_\la)^2*_e}=
\que{(q_\la)^2c^2}{2\pi *_e},$$
which is invariant, because $q_\la,\ c$ and $*_e$ are invariant. In fact the
variations of $G_e$ over time are those which are required to preserve the
invariance of the gravitational constant against the variations of $l_e$, i.e.
those of $k_\la$.

\section{Evolution of some variables and characteristics of the Universe since
t=0}

In 1929 Hubble discovered that the Universe is expanding and, according to
the ``Big Bang"\ theory, everything started with the breaking in of an
enormous amount of energy contained within a relatively small space, whose
volume has been increasing since that break-in at $t=0$. In the first basic
principle of this hypothesis we have suggested that the initial space was
configured as the 3-dimensional spherical surface $w^2+x^2+y^2+z^2=(R_i)^2$,
full of photons of very high energy, which then dispersed in all directions.

Fig. 1 shows the upper half of the 2-dimensional spherical surface
$x^2+y^2+z^2=R^2$, in which we also see a smaller circle of radius
$r=R\sin\va$. The intersection of this surface with the plane
$z=0$ is the circumference: $x^2+y^2=R^2$ situated on that plane.
On the other hand, this 2-dimensional spherical surface can be
viewed as having been generated by the differential surface
element $2\pi r  Rd\va=2\pi R^2\sin\va d\va$, by being integrated
between $\va=0$ and $\va=\pi/2$. In effect:
$$2\pi R^2\int^{\pi/2}_0\sin\va d\va=2\pi R^2,$$
which is the area of the said upper half of the spherical surface of radius
$R$.

\begin{figure}[h]
\centering
\resizebox{0.70\columnwidth}{!}{\includegraphics{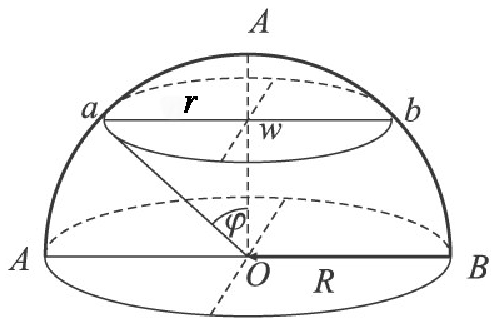}}
\caption{Fig. 1}
\end{figure}


Using the same method, we may consider that the 2-dimensional
spherical surface $x^2+y^2+z^2=R^2$ is the intersection of the
3-dimensional spherical surface $w^2+x^2+y^2+z^2=R^2$ with  the
plane surface (also 3-dimensional) $w=0$, and imagine that
3-dimensional spherical surface as having been generated by the
differential element of volume $dV=4\pi(R\sin\va)^2  R  d\va$,
whence:
$$\que V2=4\pi R^3\int^{\pi/2}_0\sin^2\va d\va=
4\pi R^3 \que\pi4=\pi^2R^3,$$
i.e.,
$$ V=2\pi^2R^3.$$

After this digression it is easier to imagine what a 3-dimensional surface is.
Because of the manner of its generation, it must possess volume, not area, and
because its points must fulfill the condition $w^2+x^2+y^2+z^2=R^2$, it must
act as a frontier between the internal points in the 4-dimensional sphere
delimited by it, in which $w^2+x^2+y^2+z^2<R^2$, and those points external to
it in which $w^2+x^2+y^2+z^2>R^2$. The link between the 4 coordinates $w,x,y$
and $z$ is analogous to that which exists between the 3 coordinates of the
2-dimensional spherical surface; this is nothing more
than a simple relation with
the curvature of the surface, which in all the spherical surfaces is the same
at all its points, because it is at all points equal to $1/R$. It follows,
from this, that the said link does not require the existence of differences
inherent to positions in space, and this is a necessary condition for the
truth of the first postulate of the Special Theory of Relativity. There are
no such differences either in the inner points nor in those external to its
boundaries delimited by $x^2+y^2+z^2=R^2$ or by $w^2+x^2+y^2+z^2=R^2$, since
both are unaffected by geometric conditions.

\vskip 6pt
\begin{center}*\ \ \ *\ \ \ *\end{center}\vskip 16pt

Zero-point radiation is inherent to space, and can therefore be considered to
be an intrinsic component of its essential nature, no matter what this may be.
According to the hypothesis formulated above, zero-point radiation has the
following characteristics:
\begin{itemize}\itemsep=-2pt
\item Its photons of greatest energy have a wavelength of
$x_tq_\la=k_u\left(\que{R_i+ct}{R_0}\right)q_\la$.
\item It is relativistically invariant. This implies that during every time
lapse $x^3t$ one photon of wavelength $x$ will strike on a given area $l^2$.
In other words, the abundance of its photons is inversely proportional to the
cubes of their wavelengths.
\item The wavelengths of its photons will increase proportionally to the
increase in the radius of the Universe $R_t=(R_i+ct)$ l.y.
\end{itemize}

The variation in the wavelengths of the photons of zero-point radiation is
similar to the variation in the lengths of the tracings which could be drawn on a
balloon made of a perfectly elastic membrane. The first tracings, at $t=0$,
when the balloon has a radius or $R_{B0}$, would increase in proportion to the
radius of the balloon $R_{Bt}$, as it inflates. The similarity is closer if
we imagine not tracings but perfectly elastic fibres which would form a
physical part of the balloon. These fibres would stretch, but would still
constitute a fixed network, while zero-point radiation is a network which is
spreading at the speed of light. However, the analogy holds, so far it
concerns the increase in lengths which is inherent to the nature of spherical
surfaces displayed by both the balloon in our example, and space, according to
the proposed hypothesis. The suggestion that space is configured as a
3-dimensional spheric surface is by no means a trivial one.

\vskip 6pt
\begin{center}*\ \ \ *\ \ \ *\end{center}\vskip 16pt

The measurements made by Sparnaay and Lamoreux allow us to infer that the
present wavelength of the photons with the greatest energy in zero-point
radiation is $x=5.257601\to10^{27}q_\la$, which implies an energy flow per
$(q_\la)^2$ during every $q_\tau$, which is given by:
$$\que{E_{0x}}{(q_\la)^2q_\tau}=\que1{(q_\la)^2q_\tau}
\que{hc}{q_\la}\left(\sum^\infty_xn^{-4}\right)$$
$$=
\que1{(q_\la)^2q_\tau} \que{2\pi}\al k_\la
\left(\sum^\infty_xn^{-4}\right)m_ec^2.$$

Since $q_\tau=q_\la/c$ is the minimum lapse of time which can apply to
electromagnetic waves, and since these waves move at the speed of light, the
amount of energy in the zero-point radiation per $(q_\la)^3$ is given by:
$$\que{E_{0x}}{(q_\la)^3}=\que1{(q_\la)^3}\que{hc}{(q_\la)}
\left(\sum^\infty_xn^{-4}\right)$$
$$=\que1{(q_\la)^3} \que{2\pi}\al
k_\la \left(\sum^\infty_xn^{-4}\right)m_ec^2,$$ and since
$$\sum^\infty_xn^{-4}=\que1{3x^2}+\que1{2x^4}+\que1{3x^5}-\que1{6x^7}+\cdots,$$
we can write as a first approximation:
$$\que{E_{0x}}{(q_\la)^3}=\que{2\pi}\al\que{k_\la}{3x^3}m_ec^2;$$
whence
$$\que{E_{0x}}{(l_e)^3}=\que{2\pi}{3\al}\que{(k_\la)^4}{x^3}m_ec^2,$$
with a relative error less than $\que1{2x^4}:\que1{3x^3}=\que3{2x}$,
which for the present value of $x$ is $\vare<2.853\ti10^{-28}$.

If we introduce the previously given
value of $x$, and $k_\la=8.143375\ti10^{20}$
and the transformation coefficient of $(l_e)^3$ to (l.y.)$^3$,
$k_{ely}=3.783997\ti10^{91}$ we obtain $E_{0x}=3.286237\ti10^{94}m_ec^2$ per
(light year)$^3$.

According to our hypothesis the radius of the Universe expressed in light
years is very approximately equal to its age expressed in years, because
$R_u=R_i+ct$ and $R_i/R_u$ must be insignificant.

The age of the Universe is estimated through the Hubble constant, $H_u$, whose
value has been estimated as not being greater than
$\que{90\ {\rm km/s}}{3.26\ti10^{6}{\rm l.y.}}$ and not less than
$\que{60\ {\rm km/s}}{3.26\ti10^{6}{\rm l.y.}}$. Its median value,
$\que{70\ {\rm km/s}}{3.26\ti10^{6}{\rm l.y.}}$ corresponds
to an age equal to
$1.39\ti10^{10}$ years and to the radius $R_u=1.39\ti10^{10}$ l.y. For this
value, the total energy of the zero-point radiation in the Universe is given
by:
$$E^*_U=2\pi^2(1.39\ti10^{10})^3 3.286237\ti10^{94}m_ec^2$$
$$=
1.742\ti10^{126}m_ec^2.$$

This value is immensely greater than the energy equivalent of the mass of the
Universe as estimated in [5], pg. 2, which is $1.55\ti10^{79}m_ec^2$.

It is obvious that, if the radius of the Universe at $t=0$ was $R_i\ll R_u$,
and if the wavelength of the photons with greatest energy in zero-point
radiation has not varied, the expansion of the Universe would have required
the increasing inflow of an enormous flow of photons, in order to keep
unchanged the amount of energy in the form of zero-point radiation per unit of
volume, in a volume $2\pi^2(R_i+ct)^3$ which grows in proportion to the cube
of the radius of the Universe, $R_t=R_i+ct$.

The density of the energy of zero-point radiation is given by:
$$\que{E_{0x}}{(q_\la)^3}=\que{hc}{(q_\la)^4}\sum^\infty_xn^{-4} $$
$$=
\que{hc}{(q_\la)^4}\left(\que1{3x^3}+\que1{2x^4}+\que1{3x^5}-
\que1{6x^7}+\cdots\right).\eqno{(7)}$$

The volume of the Universe is given by:
$$V_{ut}=2\pi^2(R_t)^3(q_\la)^3=2\pi^2(R_i+ct)^3(q_\la)^3,$$
where $R_i$ is the radius of the Universe at $t=0$ and $t$ is the time,
expressed in $q_\tau$, elapsed since $t=0$.

The total energy of the Universe after the lapse of $t  q_\tau$
since $t=0$ is, therefore:
$$E^*_U=2\pi^2(R_i+ct)^3(q_\la)^3 \que{hc}{(q_\la)^4}$$
$$ \times \left\{
\que1{3x^3}+\que1{2x^4}+\que1{3x^5}-\que1{2x^7}+\cdots\right\}=$$
$$=\que{2\pi^2hc}{q_\la}\left\{\que{(R_i+ct)^3}{3x^3}+ \right .
$$
$$ \left .
\que{(R_i+ct)^3}{2x^4}+\que{(R_i+ct)^2}{3x^3}-
\que{(R_i+ct)^3}{6x^7}+\cdots\right\}.$$

If the wavelength of the photon of greatest energy in zero-point radiation is
proportional to the radius of the Universe, its value when that radius
measured $R_iq_\la$ must have been $(R_i/k)q_\la$, and when the radius
measures $(R_i+ct)q_\la$ the wavelength must be $(R_i+ct)q_\la/k_u$, where
$k_u$ is a natural number, whence:
$$E^*_U=\que{2\pi^2hc}{q_\la}  k_u^3\left\{\que13+\que1{2x}+\que1{3x^2}-
\que1{6x^4}+\cdots\right\}.\eqno{(8)}$$
The terms within the bracket after the first term are insignificant in
comparison to it when $x$ is very large, and form a remainder what could
represent the energy needed to maintain the curvature of the Universe. We
should remember here that the condition for the relativistic invariance of the
spectrum of zero-point radiation is precisely that which causes
its total energy in the Universe
to remain substantially constant, whilst the radius of
the Universe increases indefinitely. Moreover (8) shows that the energy needed
to maintain the curvature of the Universe is inversely proportional to the
radius of the Universe, i.e. proportional to the said curvature.

Table 1 gives a list of values for $x,\ k_\la,\ \Delta_sk_\la,\ R,\ R_s$ and
$R/R_s$ as they vary with values of $z=k_\la/x$.

\vskip 6pt
\begin{table*}
\caption{Values for $x$, $k_\la$, $\Delta_xk_s$, $R$, $R_s$ and $R/R_s$ for some values
of $z=k_\la/x$. For $z\ra\al/2\pi=1.161410\ti10^{-3}[B]_m\ra\infty;$\ \ $x\ra0$;\ \
$k_\la\ra0$. There cannot exist real values of $k_\la<1$, or
values of $x<2\pi/\al=861$. The values of $z$ in lines 6-12
correspond to some of the values of $R_0/R_s$ in Table 3.
$R_0$ is the length of the radius of the Universe at the moment when the last
elementary particle was generated from the initial photons.}
\begin{ruledtabular}
\begin{tabular}{ccccccc}
$z=k_\la/x$ & $x\ (q_\la)$ & $k_\la\ (q_\la)$ & $\Delta_s
k_\la\left(\que{q_\la}s\right)$ \footnotemark[1] & $R_0$ (l.y.) &
$R_s$ (l.y.) &
$R_0/R_s$\\
\hline
$1.15\ti10^{-3}$ & $6.0745\ti10^8$ & $6.9857\ti10^5$ &
$1.034\ti10^7$ &
$1.606\ti10^{-9}$ & $1.292\ti10^{21}$ & $1.243\ti10^{-30}$\\
$10^{-4}$ & $5.299764\ti10^{13}$ & $5.299764\ti10^9$ &
$4.495\ti10^6$ &
$1.401\ti10^{-4}$ & $1.703\ti10^{17}$ & $8.227\ti10^{-25}$\\
$10^{-5}$ & $4.746727\ti10^{18}$ & $4.746727\ti10^{13}$ &
$8.990\ti10^4$ &
$12.55$ & $1.901\ti10^{13}$ & $6.602\ti10^{-13}$\\
$10^{-6}$ & $4.691894\ti10^{23}$ & $4.691894\ti10^{17}$ &
$8.990\ti10^3$ &
$1.240\ti10^6$ & $1.923\ti10^9$ & $6.448\ti10^{-4}$\\
$5\ti10^{-7}$ & $1.500432\ti10^{25}$ & $7.502160\ti10^{18}$ &
$4.495\ti10^3$ &
$3.967\ti10^7$ & $1.203\ti10^8$ & $0.3297$\\
$4.42\ti10^{-7}$ & $2.779216\ti10^{25}$ & $1.228413\ti10^{19}$ &
$3.971\ti10^3$ & $7.348\ti10^7$ & $7.345\ti10^7$ & $1.0003$\\
$4.0098\ti10^{-7}$ & $4.522692\ti10^{25}$ & $1.813509\ti10^{19}$ &
$3.605\ti10^3$ & $1.196\ti10^8$ & $4.975\ti10^7$ & $2.404$\\
$3.9115\ti10^{-7}$ & $5.120216\ti10^{25}$ & $2.002773\ti10^{19}$ &
$3.516\ti10^3$ & $1.354\ti10^8$ & $4.505\ti10^7$ & $3.005$\\
$3.7885\ti10^{-7}$ & $6.007055\ti10^{25}$ & $2.275773\ti10^{19}$ &
$3.406\ti10^3$ & $1.588\ti10^8$ & $3.965\ti10^7$ & $4.005$\\
$3.4218\ti10^{-7}$ & $9.993193\ti10^{25}$ & $3.419470\ti10^{19}$ &
$3.076\ti10^3$ & $2.642\ti10^8$ & $2.638\ti10^7$ & $10.015$\\
$3.1681\ti10^{-7}$ & $1.468820\ti10^{26}$ & $4.653369\ti10^{19}$ &
$2.847\ti10^3$ & $3.882\ti10^8$ & $1.939\ti10^7$ & $20.026$\\
$2.9332\ti10^{-7}$ & $2.158944\ti10^{26}$ & $6.332615\ti10^{19}$ &
$2.637\ti10^3$ & $5.708\ti10^8$ & $1.425\ti10^7$ & $40.056$\\
$2.0\ti10^{-7}$ & $1.464950\ti10^{27}$ & $2.929390\ti10^{20}$ &
$2.397\ti10^3$ &
$3.873\ti10^9$ & $3.080\ti10^6$ & $1.257$\\
$1.75\ti10^{-7}$ & $2.855570\ti10^{27}$ & $4.997248\ti10^{20}$ &
$2.098\ti10^3$ & $7.550\ti10^9$ & $1.805\ti10^6$ & $4.183$\\
$1.549\ti10^{-7}\footnotemark[2]$ & $5.257602\ti10^{27}$ &
$8.143376\ti10^{20}$ &
$1.856\ti10^3$ & $1.390\ti10^{10}$ & $1.108\ti10^6$ & $12.451$\\
\end{tabular}
\end{ruledtabular}
\footnotetext[1]{$\Delta_ sk_\la$ is the increase  in $k_\la$ per
second}
\footnotetext[2]{Present values}
\end{table*}

\noindent

\begin{center}$-\ \ /\ \ -\ \ /\ \ -$\end{center}\vskip 16pt

The values of $x$ have been obtained through the equation:
$$x=\que{3\al}{4\pi^3z^4[B]_m},\eqno{(9)}$$
which results from dividing the two terms of equation (3) by $x^4$, taking
$r_x=1$ and $\que{k_\la}x=z$. The values of $k_\la$ derive from $k_\la=zx$,
while those of $\Delta_sk_\la$, which is the increase of $k_\la$ per second
have been obtained from those of $\Delta_sx$ through the equation
$$\Delta_sk_\la=\que34\left(\que\al{4\pi^3[B]_m}\right)^{1/4}
\que{\Delta_sx}{x^{1/4}}.$$

The values of $\De_sx$, increase of $x$ per second, are obtained by dividing
the present value of $x$, $x=5.257601\ti10^{27}q_\la$, by the present age of
the Universe expressed in seconds; $t=4.386413\ti19^{17}$ s.; this produces a
constant value of $1.198610\ti10^{10}q_\la/s$. This value seems at first sight
very large, but is in reality very small, being equivalent to
$4.147682\ti10^{-24}$~cm/s, and to $2.28\ti10^{-18}x$. If we go back in time
65 millions of years, i.e. to the end of the era of the dinosaurs, the value of
$x$ will have diminished some $2.45\ti10^{26}q_\la$ which are
only a 4.67\% of its
present value. Finally, the values for $R_s$ have been found through
$R_s=\que{M_uG}{c^2}$, which in the $(e,m_e,c)$ system is written as:
$$R_s=M_0m_e  G_e(l_ec^2m_e^{-1})=\que{M_0}{2\pi(k_\la)^2}l_e=
\que{M_0}{2\pi(k_\la)}q_\la$$

We deduced above that the value of the gravitational constant, $G$, remains
invariant when $x$ varies, and that the variation in its numerical coefficient
in the $(e,m_e,c)$ system is precisely the variation required for $G$ to
remain invariant against variation in the length of the radius of the electron
$r_e=l_e=k_\la q_\la$, and consequently against variation in the mass of the
electron, since the product of radius and mass, $m_xl_x=e^2/c^2$, must remain
constant.

Since the value of $R_0$ is the measurement of the radius of the Universe at
the moment when the last elementary particle was generated, values of $R_0$
where $R_0/R_s<1$ would cause its collapse and transformation into a black
hole. We can see from Table 1 that the value of $x$ which corresponds to
$R_0/R_s=1.0003$ is $2.7792\ti10^{25}$, for which the value of
$\que1{2x}+\que1{3x^2}-\que1{6x^4}$ is less than $1.8\ti10^{-26}$, and the
difference between the total energy of zero-point radiation in the Universe
as the product of the Universe's volume and the density of the energy of that
radiation, and the product which results from disregarding the said remainder
is, relatively, insignificant.

Table 1 provides a basis of reference for the estimate that the age of the
Universe is $1.39\ti10^{10}$ years and that the mass of the Universe is
$1.55\ti10^{79}m_e$, [4] pp. 2-3. The first of these estimates derives from
taking the value of the Hubble constant to be 70 km/s per megaparsec, so that
the time needed to reach the ``horizon of visibility"\ formed of luminous
bodies moving away from us at speeds very near that the light will be:
$$\que{2.997925\ti10^5\ {\rm km/s}}{70\ {\rm km/s}} 3.26\ti10^6\
\hbox{years}=1.39\ti10^{10}\ \hbox{years}$$

\vskip 6pt
\begin{center}*\ \ \ *\ \ \ *\end{center}\vskip 16pt

The intersection of a 3-dimensional spherical surface of radius $R$ with a
2-dimensional plane which passes through its centre, is a circumference with a
radius $R_m$ and a centre at the centre of that surface. Fig. 2 is a
representation of such an intersection, in which  $A$ represents the position
of an observer, $B$ is that of the ``horizon of visibility"\ up to which the said
observer can see luminous objects and $\va$ is the angle $\widehat{A\om B}$.

\begin{figure}[h]
\centering
\resizebox{0.70\columnwidth}{!}{\includegraphics{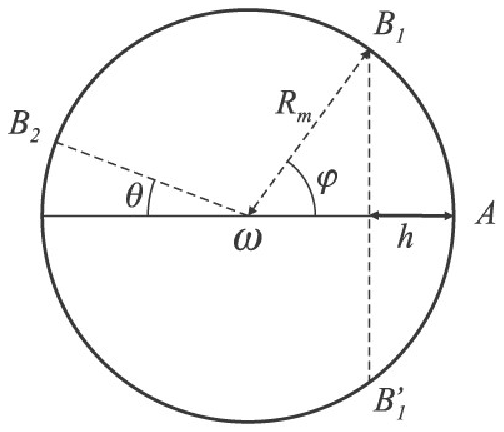}}
\caption{Fig. 2}
\end{figure}


The length of the arc $\widehat{AB_1}$ is $R_m\va$, where $R_m$ is the radius
of the visible Universe; i.e. the material Universe to which belong the
luminous objets that can be observed from $A$, and $\va$ is the angle
$\widehat{A\om B_1}$ expressed in radians.

The length of $R_m$ must be smaller than that of $R_u$, the radius of the
Uni\-ver\-se, which as a result of the dispersal of zero-point radiation in all
directions of space, measures $R_i+ct$; i.e. the initial radius $R_i$ plus as
many light years as years have elapsed since $t=0$. Therefore $R_m\va=R_t$,
where $\va>1$ and $R_t=R_i+ct$.

On a 2-dimensional spherical surface, the observer at $A$ an only see those
luminous objects which  are situated on the surface of the spherical zone of
height $h=R_m(1-\cos\va)$, whose area, $2\pi R^2(1-\cos\va)$, makes up a
fraction $\que{1-\cos\va}2$ of the area of the sphere $4\pi R^2_m$. Keeping in
mind our deduction of the volume of the 3-dimensional spheric surface as
$$V/2=4\pi R^3\disp\int^{\pi/2}_0\sin^2\va d\va=\pi^2 R^3,$$
whence $V=2\pi^2 R^3$, we can easily find the volume of the zone which
corresponds to the angle $\va<\pi/2$, through
$$V_z=4\pi R^3\int^\va_0\sin^2\va d\va=2\pi R^3(\va-\sin\va\cos\va),$$
which makes up a fraction
$$\que{\va-\sin\va\cos\va}\pi=\que{\va-(1/2)\sin2\va}\pi\quad\hbox{of}\ \ V.$$

The existence of a ``horizon of visibility"\ implies that we have not taken
into account those masses which may exist beyond it. The presumed basic
uniformity of the Universe allows us to include such masses by multiplying the
estimate obtained by considering the observable Universe, by
$\que\pi{\va-(1/2)\sin2\va}$, the inverse of the previous fraction. This value
is 2 for $\va=\pi/2$ and 1 for $\va=\pi$, when the
observer perceives the whole of it.

We must note that for $\va>\pi$, the observer at $A$ would double-count the
cosmic objects situated at $\pi+\va$ and would infer a length of $R_m$ that
would be greater than its true length. This possibility would be rejected,
were it not the case that images of very distant cosmic objects have been
detected which are very much alike, by looking in diametrally opposite
directions. It seems important to conduct a program to search for such
objects, which may have escaped detection by observers who were looking for
other things. Obviously the success of this search would demonstrate that the
Universe is configured as a 3-dimensional spherical surface, but any
unsuccessful search would not demonstrate the contrary, because there would be
no double-count if $\va<\pi$.

Table 2 lists possible present values for some characteristics of the Universe
which are related to the angle $\va$. In the second column are shown values
for the present speed of increase of the radius of $R_m$ given by
$V_m/c=1/\va$. In the third column appear the values for the kinetic energy of
an electron moving at a velocity $V_m$; in the last column are values for the
relation between the total mass of the Universe, $M_u$, and the mass, $M_u$,
perceived by an observer whose ``horizon of visibility"\ is that which
corresponds to the angle $\va$, $M_\va$. However, no mass beyond the ``horizon
of visibility"\ from $A$ would attract gravitationally any mass at $A$.

\begin{table}
\caption{Present values for some characteristics related to $\va$}
\begin{ruledtabular}
\begin{tabular}{cccc}
$\va$ (rad) & $\frac{V_m}{c}=\frac{1}{\va}$ & $\frac{E_{m_e}}{m_ec^2}=
\frac{1}{2\va^2}$ &
$\frac{M_u}{M_\va}= \frac{\pi}{\va- \frac{1}{2} \sin 2 \va}$\\
\hline
1.00 & 1.00 & 0.500 & 5.7607\\
1.20 & 5/6 & 0.347 & 3.634\\
1.40 & 0.7143 & 0.255 & 2.549\\
$\pi/2$ & 0.6366 & 0.203 & 2.000\\
1.75 & 4/7 & 0.164 & 1.632\\
2.00 & 1/2 & 0.125 & 1.321\\
2.25 & 4/9 & 0.099 & 1.147\\
2.45 & 0.4081 & 0.083 & 1.068\\
2.75 & 4/11 & 0.066 & 1.013\\
$\pi$ & 0.3183 & 0.051 & 1.000\\
\end{tabular}
\end{ruledtabular}
\end{table}

The effect of increasing the mass of the Universe by the factor $M_u/M_\va$
applies only to the variables $R_s$ and $R_0/R_s$ in Table 1. The estimate of
the radius of the Universe $R_u$ has been obtained from its age and is
independent of $\va$. In their turn, the values of $R_0$ are none other than
mere hypotheses of the length of the radius of the Universe at the moment at
which there ended the process of formation of matter from the initial photons
unconnected with zero-point radiation.

In [5] p. 4, the analysis of the kinetic energy of the last electron generated
from the initial photons unconnected with zero-point radiation led us to the
following equation
$$\que{E_t}{m_ec^2}=\que12\left(1-\que{R_s}{R_0}+\que{R_s}{R_0+R_t}\right);
\eqno{(10)}$$
where $E_t$ is the kinetic energy of the said last electron, when the time $t$
has elapsed since $t=0$, when it was generated; $R_0$ is the length of the
radius of the Universe at $t=0$, $R_s$ is the length of the Schwarzschild
radius, and $R_0+R_t$ is that of the radius of the Universe at the moment $t$.
This equation derives from our having considered that the electron moved away
from $\om$ at a velocity close to that of light, and was subjected only to the
gravitational attraction of the rest of the mass of the Universe, which would
not significantly differ from $M_0m_e$.

From (10) we deduce that the condition for the cancelling-out of the energy of
that electron is:
$$R_t=\que{(R_0)^2}{R_s-R_0};\eqno{(11)}$$
when $R_0=R_s$ it will cancel out at $R_t\ra\infty$, while if $R_0>R_s$ it can
never cancel out.

For the evolution of the velocity of the said electron, we have deduced in [5]
pp. 4-5, the equation
$$\que{dR_m}{dt}=\left(1-\que{R_s}{R_0}+\que{R_s}{R_0+R_t}\right)^{1/2},
\eqno{(12)}$$
which implies that for $\que{R_s}{R_0}\leq1\qquad\ \que{dR_m}{dt}$ cannot
cancel out, and that when $R_t\ra\infty$ the value of $\que{dR_m}{dt}$ tends
towards $\left(1-\que{R_s}{R_0}\right)^{1/2}$.

Finally, analysis of the speed at which the horizon of visibility is receding
$$\que{d\widehat{AB}}{dt}=\va\que{dR_m}{dt}+R_m\que{d\va}{dt},$$
leads us to infer that for $t=0$, $\va=0$ and for $t\ra\infty$,
$\va\ra(1-R_s/R_0)^{-1/2}$, [5] pp. 8-9. We can definitely say
that both the rate at which the radius of the material Universe is now
increasing and the present magnitude of the angle $\va$ which defines the
``horizon of visibility", are determined by the value of $R_s/R_0$, that is by
the relation between the Schwarzschild radius and the radius of the Universe
at the moment $t=t_0$ when the last electron was generated from the initial
photons unconnected to zero-point radiation. Table 3 shows the present values
of the angle $\va$, the rate of increase in the radius of the material
Universe, $dR_m/dt$, and the relation between the mass of the Universe, $M_u$
and that of the visible Universe $M_\va$. These values are given for the
values of $R_0/R_s$ falling between 1.000 and 628 which appear in Table 1, and
for values which in Table 2 are equal to $\pi$ radians, 2.45 radians, 2
radians, $\pi/2$ radians and 1 radian. This completes a fair panorama of the
effects of the possible values of $R_0/R_s$ on important characteristics of
the Universe.

\vskip 6pt
\begin{table}
\caption{Present values of the rate of increase in the radius of the material Universe,
$dR_m/dt$, of the angle $\va$, and of the relation $M_u/M_\va$ for some values
of $R_0/R_s$}
\begin{ruledtabular}
\begin{tabular}{cccc}
$\frac{R_0}{R_s}$ & $\va$ (rad) & $\frac{dR}{dt}c$ &
$\frac{M_u}{M_\va} = \frac{\pi}{\va- \frac{1}{2}\sin2\va}$\\
\hline
$<1$ & negative & negative & --\\
1.0044 & 15.11 & 0.066 & 0.2017\\
1.11274 & $\pi$ & 0.3183 & 1.0000\\
1.20 & 2.45 & 0.408 & 1.0681\\
1.333 & 2.00 & 0.500 & 1.3209\\
1.50 & 1.73 & 0.577 & 1.6653\\
1.682 & $\pi/2$ & 0.637 & 2.0000\\
3.273 & 1.200 & 0.833 & 3.6434\\
5.00 & 1.118 & 0.894 & 4.3356\\
10.00 & 1.0541 & 0.949 & 5.0299\\
20.00 & 1.0260 & 0.975 & 5.3907\\
628 & 1.0008 & 0.9992& 5.7487\\
\end{tabular}
\end{ruledtabular}
\end{table}

\begin{itemize}\itemsep=-2pt
\item For $R_0/R_s=1$, $\va\ra\infty$, $dR/dt\ra0$: This value of $\va$ is
inadmissible, as we would be contemplating an infinite series of images of he
same cosmic objects.
\item For $R_s/R_0<1$, we would obtain negative values for $dR/dt$. This does
not seem possible for $t=1.39\ti10^{10}$ years.
\end{itemize}

\section{Analysis of the Possibility of the  Big Crunch and Evaluation
of the Radius of the Material Universe}

Equation (10) in [5], which expresses the evolution of $R_m$ as a function
of $t$, and equation (14) in [5], which expresses the evolution of $\va$ as a
function of $t$, include the expression $(1-R_s/R_0)^{1/2}$, which gives a
complex number for $R_s/R_0>1$; which may mean that it is impossible for $R_0$
to be smaller than $R_s$.

If we abstract from these equations which derive from equation (5) in [5], and
consider only this equation (5), we obtain:
$$\que{dR_m}{dt}=\left(1-\que{R_s}{R_0}+\que{R_s}{R_0+t}\right)^{1/2}=
\left(1-\que{R_s}{R_0}+\que{R_s}{R_m}\right)^{1/2},$$
where $R_m$ is the present length of the radius of the material Universe. If
we substitute $\que{R_s}{R_0}$ for $(1+x)$, we obtain
$\que{dR_m}{dt}=\left(\que{R_s}{R_m}-x\right)^{1/2},$
which will be real for $R_s/R_m>x$.

The redshift in the light which comes from the very distant galaxies allows
us to know that the radius of the material Universe, $R_m$, is increasing,
that is, that $dR_m/dt>0$, which means that $R_m>R_s$. Therefore $x<1$.

We can see from Table 3 that for $R_0/R_s<1.11274$ the angle $\va$ measures
more than $\pi$ radians, which determines that $M_u/M_\va<1$. Therefore
$M_u<M_0$.

Table 3 shows that for $R_s/R_0\ra1, \ \va\ra\infty,\ \ \que{dR_m}{dt}\ra0$, and
$\que{M_u}{M_\va}\ra0$. This means that the suggested values
$\que{R_s}{R_0}=1+x$, $x<1$ are incompatible with the hypothesis which is
analysed in this paper. In other words this hypothesis is incompatible with
the ``Big Crunch".

\vskip 6pt
\begin{center}*\ \ \ *\ \ \ *\end{center}\vskip 16pt

Equation (12) can be written as
$\que{dR_m}{\left(1-\que{R_s}{R_0}+\que{R_s}{R_m}\right)^{1/2}}=dt.$

If we integrate between $t=0$, for which value $R_m=R_0$, and $t$, whose value
governs $R_m$, equation (12) becomes:
$$\int^{R_t}_{R_0}\que{dR_m}{\left(1-\que{R_s}{R_0}+\que{R_s}{R_m}\right)^{1/2}}=t$$

To solve this integration we substitute $R_m=\que{R_s}y$,
$dR_m=-\que{R_sdy}{y^2}$. Also, we simplify by
$1-\que{R_s}{R_0}=a$, so that we
have:
$$-R_s\int^{R_s/R_m}_{R_0}\que{dy}{y^2(1+ay)^{1/2}}=
R_s\left[\que{(a+y)^{1/2}}{ay}\right]^{R_s/R_m}_{R_s/R_0}+$$
$$+\que{R_s}{2a\sqrt{a}}\log\left[\que{(a+1)^{1/2}-a^{1/2}}{(a+y)^{1/2}+a^{1/2}}
\right]^{R_s/R_m}_{R_s/R_0},$$
so that if $(1-R_s/R_0+R_s/R_m)=(a+R_s/R_m)=b;$ we finally arrive at:
$$R_m=t\left(\que ab\right)^{1/2}+\que{R_0}{b^{1/2}}+\que{R_sb^{1/2}}{2a}
\log\left[\que{b^{1/2}-a^{1/2}}{1-a^{1/2}}
\que{1+a^{1/2}}{b^{1/2}+a^{1/2}}\right]\eqno{(13)}$$

Since $b=1-R_s/R_0+R_s/R_m$, equation (13) is extremely complex and to find
the value of $R_m$ we have to resort to successive iterations starting from
$R_m=\left(1-\que{R_s}{R_0}\right)^{1/2}t$, where
$\left(1-\que{R_s}{R_0}\right)^{1/2}$ is an estimate for the present value of
$\que{dR_m}{dt}$, which leads to a value for $R_m$ which is smaller than that
given by equation (13).

Table 2 shows the values towards which $v_m=dR_m/dt$ will tend for
$t\ra\infty$, $R_m\ra\infty$. Obviously, the present shorter estimate of the
length of $R_m$ must correspond to the lower of these values, $v_m=0.3183c$,
which is the value given for $\va=\pi$. For this value the visible mass
$M_\va$ would be exactly equal to $M_u$, that is the visible mass of the
Universe.

If we introduce into (13) the value $R_m=0.3183c\ti1.39\ti10^{10}$ years
$=4.42437\ti10^9$ light-years, evidently smaller than the present value of
$R_m$, after 10 interactions we arrive at $R_m=9.92345\ti10^9$ light-years.
Therefore $9.923\ti10^9$ light-years $<R_m<1.39\ti10^{10}$ light-years (14).

Within this interval fall all the possible values of the length of the radius
of the material Universe, $R_m$, which correspond to the values of $v_m$ in
Table 2 that fall between $v_m=c$, for $\va=1$ radian, and $v_m=0.3183c$, for
$\va=\pi$ radians.

\vskip 6pt
\begin{center}*\ \ \ *\ \ \ *\end{center}\vskip 16pt

Equation (8) allows us to suggest that the total energy inherent to the
curvature of the Universe can be given by:
$$E^*_{cu}=\que{2\pi^2hc}{q_\la}k^3_u\left\{\que1{2x}+\que1{3x^2}-
\que1{6x^4}+\cdots\right\},$$
where $k$ is a constant. The wavelength, $x$, of the most energetic photons in
zero-point radiation is proportional to the length of the radius of the
Universe, $R_u$, and the values of $x$ in Table 1, $x>6.07\ti10^8$, imply that
the value of the sum $\que1{3x^2}-\que1{6x^4}+\cdots$ is insignificant when
compared to $1/2x$, and we can write:
$$E^*_{cu}=\que{\pi^2hck^3}{xq_\la},$$
which states that the total energy inherent to the curvature of the space
is inversely proportional to $x$; i.e. inversely proportional to the radius of
the Universe and, therefore, directly proportional to the curvature of the
Universe.

Within a space configured as a 3-dimensional spherical surface, all points are
affected by its curvature, and from equation (7) we can obtain
$$\que{E_{cu}}{(q_\la)^3}=\que{E_{0x}}{(q_\la)^3}-\que{hc}{(q_\la)^4}
\que1{3x^3}$$
$$=\que{hc}{(q_\la)^4}\left\{\que1{2x^4}+\que1{3x^5}-\que1{6x^7}+
\cdots\right\}$$
which, given the magnitude of the possible values of $x$, can be simplified to
$$\que{E_{cu}}{(q_\la)^3}=\que{hc}{(q_\la)^4} \que1{2x^4};\qquad
\hbox{whence}$$
$$\que{E_{cu}}{(l_e)^3}=\que\pi\al\que{(k_\la)^4}{x^4}m_ec^2,$$
which expresses the energy per $(l_e)^3$ inherent to the curvature of the
Universe. This equation may be written:
$$\que{E_{cu}}{(l_e)^3}=\que\pi\al\,(z)^4\que{m_ec^2}{(l_e)^3};\qquad
\hbox{where}\ \ z=\left(\que{k_\la}x\right)^4.\eqno{(15)}$$

The increase in the radius of the Universe presupposes a proportional increase
in $x$, which is equivalent to $1.126651\ti10^{-13}q_\la/t_e$, and causes a
decrease in $E_{cu}$. This decrease produces a certain flow of energy per
unit of volume, which translates into an expansive force. In other words, the
curvature of the Universe gives up energy as it stretches, which can be
understood better with the help of the image of an arrow impelled by the
energy yielded by the distending bow as it is released.

Equation (9) can be written in the form:
$$z^4=\que{3\al}{4\pi^3x[B]_m},$$
whence
$$\que{\pa(z^4)}{\pa x}dx=\que{3\al}{4\pi^3[B]_m} \que{dx}{x^2}.$$
Therefore:
$$\que{\De E_{cu}}{(l_e)^3}m_ec^2=\que{-3}{4\pi^2[B]_m}
\que{\De x}{x^2}\que{m_ec^2}{(l_e)^3}\eqno{(16)}$$

Within space configured as a 3-dimensional spherical surface, the volume of
the electron is $2\pi^2l_e^3$. Therefore the stretching of the curvature of
space implies, for the electron, a centrifugal force given by
$$f_{ce}=2\pi^2\left(\que{-3}{4\pi^2[B]_m}\que{\De x}{x^2}\right)
m_el_et_e^{-2}.$$
The value of $\De x$ expressed in $l_e$ per $t_e$ is $\De
x=1.383518\ti10^{-34}l_e/t_e$. According to our hypothesis this value does not
change. Therefore we can write
$$f_{ce}=-\que{2.075277\ti10^{-34}}{[B]_mx^2}m_el_et_e^{-2}.\eqno{(17)}$$
For the present values $x=6.456292\ti10^6l_e$, $[B]_m=1.944468\ti10^{-5}$, we
obtain $f_{ce}=-2.560407\ti10^{-43}m_el_et_e^{-2}$. Against this centrifugal
force, the attraction determined by the mass of the Universe is:
$$f_g=\que{m_x\ti1.55\ti10^{79}m_x}{(R_{ux})^2(l_x)^2}
\que{(q_\la)^2c^2}{2\pi*}=
\que{1.55\ti10^{79}(m_x)^2c^2}{2\pi(R_{ux})^2(k_{\la x})^2m_el_e},$$
where $R_{ux}$ and $k_{\la x}$ are, respectively, the length of the radius of
the Universe expressed in $l_x$ and the length of the radius of the electron
expressed in $q_\la$, at $t_x$.

The relations $l_x=k_{\la x}q_\la$; $l_e=k_\la q_\la$; $m_el_e=m_xl_x$ allow
us to write $m_x=m_ek_\la/k_{\la x}$. By introducing this in $f_g$, we obtain:
$$f_g=\que{1.55\ti10^{79}(k_\la)^2}{2\pi(R_{ux})^2(k_{\la x})^4}m_el_et_e^{-2}=
\que{1.027876\ti10^{121}}{2\pi(R_{ux})^2(k_{\la x})^4}m_el_et_e^{-2}.
\eqno{(18)}$$

For the present values $R_u=4.666577\ti10^{40}l_e$;
$k_{\la x}=k_\la=8.143\ti10^{20}$ we obtain
$f_g=1.708229\ti10^{-45}m_el_et_e^{-2}$, which is equal to
$6.67\ti10^{-3}f_{ce}$. For
the lower limit of $R_u$, $9.92\ti10^9$ l.y. instead of $1.39\ti10^{10}$ l.y.,
$f_g=3.354\ti10^{-44}m_el_et_e^{-2}$, also very inferior to $f_{ce}$.

From (17) and (18), we obtain:
$$\que{f_{ce}}{f_g}=\que{2.075277\ti10^{-34}/[B]_mx^2}{1.027876\ti10^{121}/
2\pi(R_{ux})^2(k_{\la x})^4}$$
$$=\que{1.267482\ti10^{-154}(R_{ux}/x)^2(k_{\la
x})^4}{[B]_m}$$

According to our hypothesis $R_{ux}/x$ is constant and its value is
$7.227951\ti10^{33}$. Therefore:
$$\que{f_{ce}}{f_g}=\que{6.621742\ti10^{-87}[k_{\la x}]^4}{[B]_m}
\eqno{(19)}$$

For the present values $k_{\la x}=k_\la=8.143375\ti10^{20}$;
$[B]_m=1.944468\ti10^{-5}$, we obtain $f_{ce}/f_g=149.758$, whence
$f_g/f_{ce}=6.67\ti10^{-3}$ as before.

$[B]_m$ decreases over time. For $t_x=12.55$ years after the Big Bang,
$[B]_m=1.239543\ti10^{-3}$, and at present, $t_x=1.39\ti10^{10}$ years,
$[B]_m=1.944468\ti10^{-5}$. In Table 1 we can see that $k_{\la x}$ increases
over time and, therefore, the value of $f_{ce}/f_g$ also increases. For
$t_x=12.55$ years after the Big Bang, it was equal to $9.05\ti10^{-30}$
instead of 149.78.

When $f_{ce}=f_g$, i.e. when the centrifugal force which comes from the
decrease in the curvature of the Universe are equal to the gravitational
attraction of the whole mass of the Universe, we have:
$$\que{6.621742\ti10^{-87}[k_{\la x}]^4}{[B]_m}=1,$$
which happens at $t_x=4.36\ti10^9$ y.; $k_\la=3.234\ti10^{20}$;
$[B]_m=7.249\ti10^{-5}$; $x=1.650\ti10^{27}$.

We know that $R_m$ is increasing and that the centrifugal forces inherent
to the decrease in the Universe's curvature prevail over the gravitational
attraction and will always prevail over it; therefore the hypothesis which is
proposed in this paper excludes the Big Crunch.

The introduction of the centrifugal force inherent to the
stretching of the curvature of space was made through the addition
of the term $\que1{2x^4}$ which appears in the parenthesis of
equation (7). The fact of not including it in equation (3) implies
relative errors inferior to $\que32 \que1x$, which are
insignificant for the values of $x$ in Table 1. Therefore it is
unnecessary to change anything in that Table.

\section{Reflections of the constant $\alpha$  and on the Generation of Elementary
Particles}

In the $(e,m_e,c)$ system of units the equation which gives the energy of
the photons of wavelength $\la l_e$ is:
$$E_\la=\que{hc}{\la l_e}=\que{2\pi}\al \que1\la m_ec^2.\eqno{(20)}$$

For $\la=\que{2\pi}\al l_e$ we obtain $E_\la=m_ec^2$, which is the
energy equivalent of the mass of the electron whose radius, $r_e$,
measures $l_e$. The wavelength $2\pi/\al$ is equal to the length
of a circumference of radius $R_e=l_e/\al=r_e/\al$, which implies
a relationship of the scale $1/\al$ between the wavelength
$(2\pi/\al)l_e$ and the length of the circumference of radius $r_e$.
In all cases, an elementary particle of mass $m_x$ has a radius
$r_x=\que{m_e}{m_x}  l_e$ and the photons of wavelength
$\que{2\pi}\al r_x$ have an energy
$$E_x=\que{2\pi}\al\que{m_ec^2}{(2\pi/\al)(m_e/m_x)}=m_xc^2,$$
which is the energy equivalent to the mass $m_x$. The wavelength
$\lambda_x=\que{2\pi}\al r_x$ is equal to the length of a circumference of radius
$r_x/\al$.

The product $m_el_e=e^2/c^2$ is a quantic threshold. There are no
charges smaller than ``$e$", or speeds greater than ``$c$". On the
other hand the relationship $r_x  m_x=r_e  m_e$ is the requisite
condition for the spin of a particle of mass $m_x$ and radius
$r_x$ to be equal to $\hbar/2$, so that a photon of wavelength
$\la_x$ has an energy of $E_x=(m_e/\la_x)c^2=m_xc^2$,
equal to the energy equivalent of the mass of the particle, while
the relationship between the wavelength of the photon with energy
$E_x$ and the length of the circumference whose radius is that of
the elementary particle of mass $E_x/c^2$, is always equal to
$1/\al$.

The value of the constant $\al$ is not affected by hypothetical
changes in the wavelength ``$x$"\ of the photons of greatest
energy in zero-point radiation. However any change in ``$x$"\
would produce a change in the length of the radius of the electron
$r_e=k_\la q_\la$, which is determined by the fact that, at this
distance from its centre, the centrifugal force of the repulsion
of its charge against itself is equal to the centripetal force
inherent to the interaction of the particle with zero-point
radiation. This change in the length of its radius produces an
inversely proportional change in the mass of the electron, because
$m_x  r_x$ must continue to be equal to $m_er_e$. Therefore the
wavelength of a photon whose energy is equal to $m_xc^2$ will also
change in proportion to the change in length in the electron's
radius, so that the relationship between the new wavelength and
the new radius continues to be $1/\al$.

The expression of the equivalence between those centrifugal and centripetal
forces relates $k_\la$ and $x$ through equation (9), which is an expression of
equation (17) of [4] and can be written:
$$x=\que{3\al/4\pi^3}{z^4[B]_m},$$
where
$$z=\que{k_\la}x $$
$$[B]_m=\sum^\infty_1\left(\que{2\pi z}\al\right)^m $$
$$T_m=(-1)^{m+1}$$
$$\left\{\que1{m+1}+\que2{m+2}-\que3{m+3}-1-
\que{m(m-1)}6\right\}\que1{m+3}.$$

The series $[B]_m$ has a positive and finite value for $z<\que\al{2\pi}$ and
has no value for $z\geq\que\al{2\pi}$. On the other hand
$\que{k_{\la t}}{x_t}=z_t<\que\al{2\pi};$ where $x_tq_\la$ is the
wavelength of the most energetic photons in zero-point radiation $t$ years
after the Big Bang and $k_{\la t} q_\la$ is the radius of the electron at that time.

Because of the quantic threshold $m_xr_x=e^2/c^2$; $m_tk_{\la t}=m_el_e$, whence
$m_t=\que{m_el_e}{k_{\la t}}$, and the wavelength of the photon with energy
$m_tc^2$, expressed in $(q_\la)$, is
$$\la=\que{2\pi}\al k_{\la t}$$
whilst the wavelength of the most energetic photons in
zero-point radiation is
$$ x_t > \frac{2 \pi}{\alpha} k_{\la t} ,$$
so that there could never occur the generation of electrons or any other
elementary particle. The value of the product $\que{2\pi}\al z$ increases as
$x$ decreases and tends toward 1 as $x$ tends towards 0. For the present value
of $x$, the value of $z$ is $1.548876\ti10^{-7}$, and, for
$z=1.150\ti10^{-3}=0.99(\al/2\pi)$, $x=6.0745\ti10^8$, which corresponds to
$t=1.606\ti10^{-9}$ years after $t=0$, very approxi\-ma\-te\-ly $0.05s$ after the
Big Bang, and exposes the extremely rapid variation of $x$ between the said
value and $z=\al/2\pi$.

Some cosmologists suppose that the present laws of Physics did not apply
during the time immediately after the Big Bang and that the generation of the
elementary particles happened on a short period of time. Nevertheless it is
better to suppose that zero-point radiation cannot generate elementary
particles and that the primeval space which emerged with the Big Bang
contained other photons able to generate those particles which are now in our
Universe.

\section*{REFERENCES}

[1] L. Abbot: {\em Inv. y Ciencia}, p. 80-83 (Nov. 1983).

[2] R. Alvargonz\'alez: {\em Rev. Esp. Fis.} {\bf 14 (4)}, 32 (2000).

[3] R. Alvargonz\'alez: {\em arXiv: physics}/9311027 V1, 6 Nov 2003.

[4] R. Alvargonz\'alez: {\em arXiv: physics}/0311139 V2, 23 May 2005.

[5] R. Alvargonz\'alez and L. S. Soto: {\em arXiv: phyusics}/0408016 VI, 3 Aug
2004.

[6] T. H. Boyer: {\em Phys. Rev.} {\bf 182}, 1374 (1969).

[7] T. H. Boyer: {\em Sci. Am.} {\bf N 10}, 42 (1985).

[8] S. K. Lamoreaux: {\em Phys. Rev. Let.} {\bf 78}, 5 (1997).
\end{document}